\documentclass[
 reprint,
 floatfix,
 amsmath,amssymb,
 aps,
]{revtex4-2}

\usepackage{graphicx}% Include figure files
\usepackage{dcolumn}% Align table columns on decimal point
\usepackage{bm}
\begin{document}

%\definecolor{ultramarine}{rgb}{0.07, 0.04, 0.56}
\newcommand{\is}[1]{\textcolor{ultramarine}{\textbf{(IS: #1)}}}

\newcommand{\apjs}{ApJS}
\newcommand{\apjl}{ApJ Letters}
\newcommand{\mnras}{MNRAS}
\newcommand{\aap}{Astronomy \& Astrophysics}
\newcommand{\jcap}{Journal of Cosmology and Astro-Particle Physics}

\preprint{APS/123-QED}

\title{Fermi Bubbles Without AGN: $\gamma$-Ray Bubbles in MHD Galaxy Formation Simulations with Full Cosmic Ray Spectra}% Force line breaks with \\

\author{Isabel S. Sands}
\affiliation{TAPIR, California Institute of Technology, Mailcode 350-17, Pasadena, CA 91125, USA}
 %Lines break automatically or can be forced with \\
\author{Philip F. Hopkins}%
\affiliation{TAPIR, California Institute of Technology, Mailcode 350-17, Pasadena, CA 91125, USA}%\\
\author{Sam B. Ponnada}%
\affiliation{TAPIR, California Institute of Technology, Mailcode 350-17, Pasadena, CA 91125, USA}%

\date{\today}% It is always \today, today,
             %  but any date may be explicitly specified

\begin{abstract}
    For the first time, we show in MHD simulations with cosmological initial conditions that bi-lobed gamma-ray outflows similar to the Fermi bubbles can form from star formation and supernova feedback, without involvement from active galactic nuclei (AGN). We use simulations run with full MHD and dynamical, on-the-fly multi-species cosmic ray transport in MeV-TeV energy bins to model gamma-ray emission in Milky Way-mass spiral galaxies from neutral pion decay, relativistic non-thermal Bremsstrahlung, and inverse Compton scattering. We find that these $\gamma$-ray outflows are present in all three Milky-Way mass simulated galaxies. The amplitude, shape, and the composition of the gamma-ray spectrum of these bubbles fluctuates over time, with lepton-dominated and hadron-dominated phases. Spectra in which there is $\mathcal{O}(1)$ more $\gamma$-ray flux from inverse Compton scattering than neutral pion decay are a good fit to the measured Fermi-LAT spectrum. Additionally, these simulations predict multi-wavelength features in soft x-rays and synchrotron radio, potentially providing new observational signatures that can connect the circumgalactic medium to cosmic ray physics and activity in the galactic center. 
\end{abstract}

%\keywords{Suggested keywords}%Use showkeys class option if keyword %display desired
\maketitle

%\tableofcontents

\indent{\textbf{Introduction}}-- There is abundant observational evidence across multiple wavelengths of outflows from the central regions of galaxies \cite{sarkar2024fermierositabubbleslooknuclear}. These outflows are typically thought to arise from clustered supernovae in galaxies with active star formation, or from radiative or kinetic modes of active galactic nucleus (AGN) feedback, and have been observed in a variety of galaxies across a range of masses \cite{Veilleux_2020}. Because observables of these outflows span the electromagnetic spectrum, from diffuse radio emission to x-ray and $\gamma$-ray bubbles, such features are ideal targets to test different feedback mechanisms that can regulate star formation in galaxies.

Within the Milky Way (MW), the Fermi Large Area Telescope (LAT) has detected a pair of bi-lobed, bubble-like features in $\gamma$-ray emission above and below the plane of the MW's galactic disk, spanning approximately 10 kpc and originating from the galactic center \cite{Su_2010}. The $\gamma$-ray bubbles are accompanied by a pair of larger x-ray bubbles detected by ROSAT and eROSITA, approximately 12-14 kpc in length \cite{2020Natur.588..227P}. The origins of both the Fermi and eROSITA bubbles remain elusive: whether they are driven by AGN or star formation, a single event or multiple episodes in which energy and material are ejected from the galaxy, and whether the $\gamma$-ray emission is from hadronic or leptonic processes \cite{Yang_2018, 2022MNRAS.514.2581M, 2022NatAs...6..584Y, 2022MNRAS.516.1539O}. 

\begin{figure*}
    \centering
    \includegraphics[width=\linewidth]{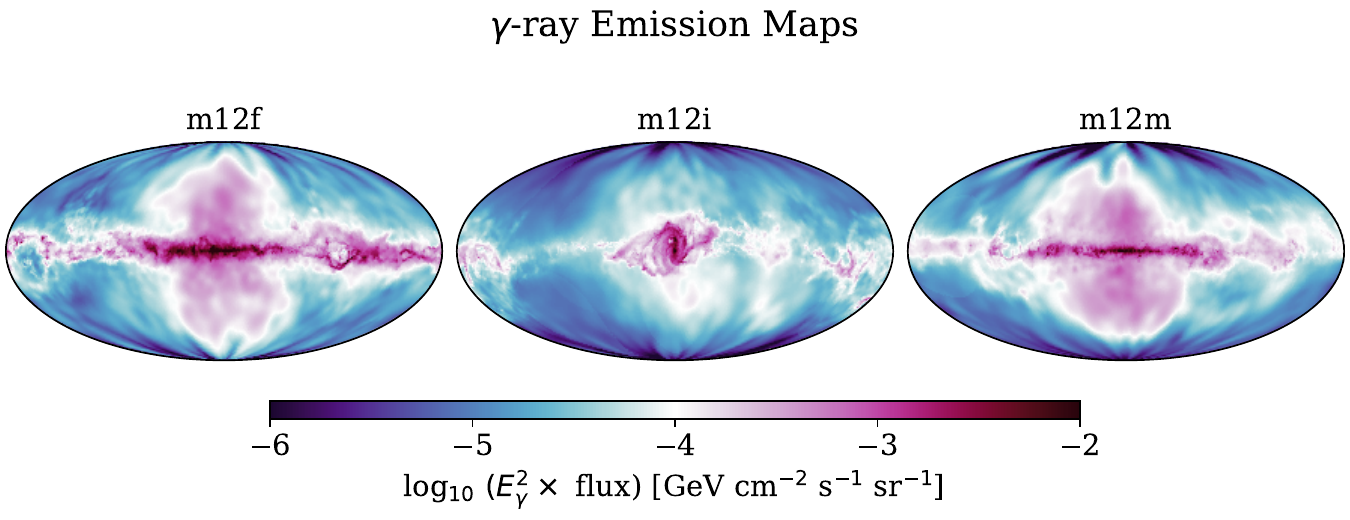}
    \includegraphics[width=\linewidth]{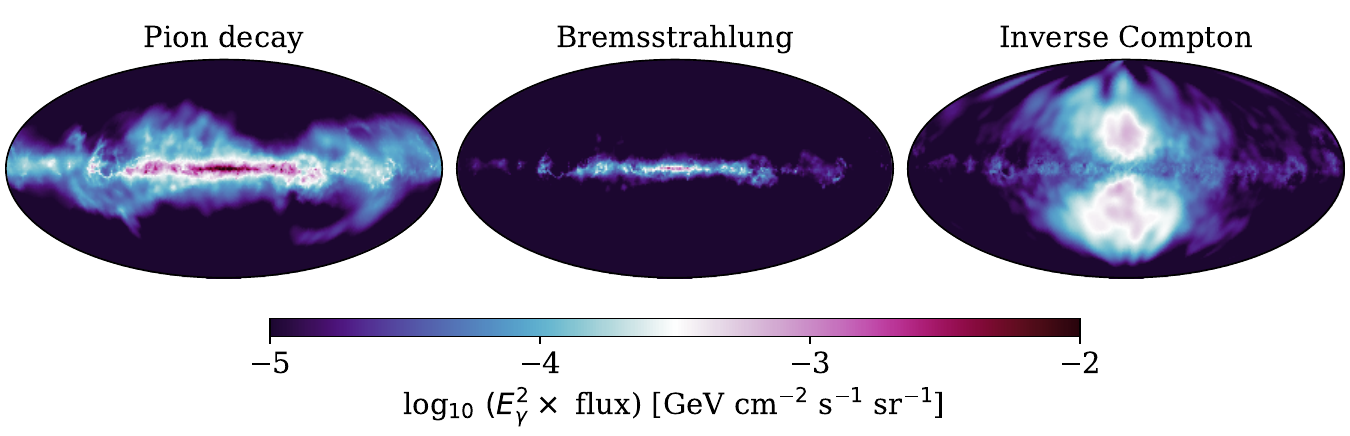}
    \caption{\textbf{Top}: total flux of 5-50 GeV $\gamma$-rays for the three simulated MW analogs in our sample. All galaxies show a distinct bi-lobed feature above and below the plane of the galactic disk; this feature is less pronounced for \textbf{m12i}, which has a warped disk and a strong magnetic field at its center, which leads to higher CR lepton loss rates than in \textbf{m12f} or \textbf{m12m}. \textbf{Bottom}:  $\gamma$-ray emission in galaxy \textbf{m12m}from the three relevant CR interactions: neutral pion decay, relativistic Bremsstrahlung, and inverse Compton scattering. ICS is responsible for most of the bubble structure, with a slight hadronic contribution at lower latitudes.  }
    \label{fig:all_bubbles}
\end{figure*}

Most past studies of the origin and nature of the Fermi bubbles have relied on fitting the Fermi-LAT data to static templates generated by cosmic ray propagation codes or observed gas maps \cite{Su_2010,2014ApJ...793...64A},  or idealized hydrodynamic simulations \cite{2012ApJ...761..185Y,2015MNRAS.453.3827S,2022NatAs...6..584Y, 2022MNRAS.514.2581M, 2022MNRAS.516.1539O}. In the case of the former, fits to static templates are inherently unable to probe the time evolution of these structures. On the other hand, idealized hydrodynamic simulations can study the growth over time of outflows fueled by a specific process (e.g., stellar winds or an AGN burst) but typically impose a galactic potential, make simplifying assumptions about magnetic field structure and the hot circumgalactic medium (CGM), and do not evolve the CRs in a cosmological setting. It has only recently become possible to simulate CRs alongside full magnetohydrodynamics in cosmological simulations \cite{Booth_2013, Salem_2013, 2016ApJ...816L..19G, Butsky_2018, Chan_2019, Buck_2020, Werhahn_2021, Pfrommer_2022, Farcy_2022, thomas2022cosmic}. Most of these simulations evolve CRs as a single fluid element in a single energy bin. Several of these simulations have shown that CR pressure allows for the formation of kpc to Mpc-scale gas outflows, indicating that CR pressure may play an important role in the formation of the Fermi and eROSITA bubbles \cite{2024MNRAS.535.1721P,2025A&A...699A.125R}. 

More recently, new numerical methods have enabled the self-consistent evolution of individual CR species in magnetohydrodynamic galaxy formation simulations run from cosmological initial conditions, allowing for detailed studies of astro-particle phenomena and the forward-modeling of multi-wavelength observables arising from CR interactions with the interstellar medium (ISM) \cite{Hopkins_CR}. In this letter, we analyze the simulations introduced in \cite{Hopkins_CR}, in which multi-species cosmic rays (electrons, positrons, protons, anti-protons, and heavier nuclei) are simulated alongside full magnetohydrodynamics. We find that, for our sample of three MW-mass galaxies, extended $\gamma$-ray bubbles and halos are a ubiquitous feature that arise from the inverse Compton scattering of CR leptons injected by supernovae and fast stellar winds. The $\gamma$-ray spectrum of these bubbles is consistent with the Fermi-LAT observations, and we further find that these $\gamma$-ray features have counterparts in x-ray and synchrotron radio emission. 

This letter is structured as follows. We first provide a brief description of the simulations and the methods used to model $\gamma$-ray emission from the simulated CRs. We then show how different particle physics interactions contribute to the $\gamma$-ray emission, creating bi-lobed bubbles. We further demonstrate how the structure and composition of these bubbles (i.e. whether the $\gamma$-ray emission is predominantly produced by hadronic or leptonic interactions) changes over million-year (Myr) time scales. Finally, we show that the galactic outflows that lead to the $\gamma$-ray bubbles produce features in other wavelengths, including synchrotron radio and soft x-ray. This work marks the first time that Fermi bubble-like $\gamma$-ray features have been produced in simulations of MW-mass galaxies evolved from cosmological initial conditions.

\textbf{Methods}: In this work, we analyze the same set of simulations first introduced in Hopkins et. al. 2022 \cite{Hopkins_CR}. These simulations are controlled re-starts of single-bin MHD-CR simulations evolved to late times, and were run with the GIZMO code and Feedback in Realistic Environments (FIRE) physics. There is feedback from type Ia and II supernovae, radiation (e.g., photo-electric and photo-ionization heating), and fast stellar winds from O/B and AGB star mass loss \cite{Hopkins_2018, Hopkins_2022}. We refer the reader to these works for details of numerical implementation, particularly \cite{Hopkins_CR} for details of the methods for CR transport. The simulations adopt a single power-law injection for the cosmic rays, with 10\% of supernova and stellar wind shock energy going into CRs, and 2\% of that into leptons. There are no black holes or AGN in these simulations; CRs are injected only from stellar sources. None of the parameters used in our simulations are tuned to match the MW or Fermi-LAT observations. We use local ISM observations from Voyager and AMS to set the relative normalizations of the cosmic ray spectra for individual species at redshift $z = 0.05$.

 For our set of three realizations of MW-like galaxies (\textbf{m12f}, \textbf{m12i}, and \textbf{m12m}), we model $\gamma$-ray emission from neutral pion decay, relativistic non-thermal Bremsstrahlung, and inverse Compton scattering (ICS) in post-processing, following the procedure described in \cite{sands2025galacticcentergammarayemission}. We extend the emission for relativistic Bremsstrahlung and ICS to the x-ray regime, so we cover photons emitted from 0.1 keV to 100 GeV. We neglect x-ray emission from non-relativistic thermal Bremsstrahlung, as there is little hot gas in the region of interest, and past analysis has demonstrated that CR pressure in the CGM can suppress thermal Bremsstrahlung emission \cite{silich2025xrayemissionsignaturesgalactic}. When calculating the $\gamma$-ray spectrum for the bubbles, we mask the central $\pm 20^{\circ}$ to exclude the galactic disk, and then select a conic region within a slope of $\pm 45^{\circ}$ in latitude vs. longitude.

\textbf{Overview of results}: All three of the galaxies in our sample form bi-lobed $\gamma$-ray bubbles (Fig. \ref{fig:all_bubbles}, top); these features are present at the beginning of the controlled restart, which indicates that cosmic ray pressure acting against thermal pressure in the CGM plays an important role in inflating the bubbles  by acting against thermal pressure and allowing cool gas to expand \cite{Ji_2020}. Throughout the simulation, CR pressure is comparable to or exceeds thermal pressure in the inner CGM. Similar to the Fermi bubbles, these features are relatively flat in surface brightness. The round ``bubble" structure arises from the inverse Compton scattering of CR leptons. $\gamma$-ray emission from $\Pi_0$ decay and relativistic Bremsstrahlung are brightest within the plane of the galactic disk, but can also contribute to the emission from the bubbles (Fig. \ref{fig:all_bubbles}, bottom). 
%Expand on this: from Ji 2020, if you have CR pressure dominate over thermal pressure, virial shock is delayed and you don't get hot halo. Cool gas is volume-filling. 

\begin{figure}[h!]
    \centering
    \includegraphics[width=\columnwidth]{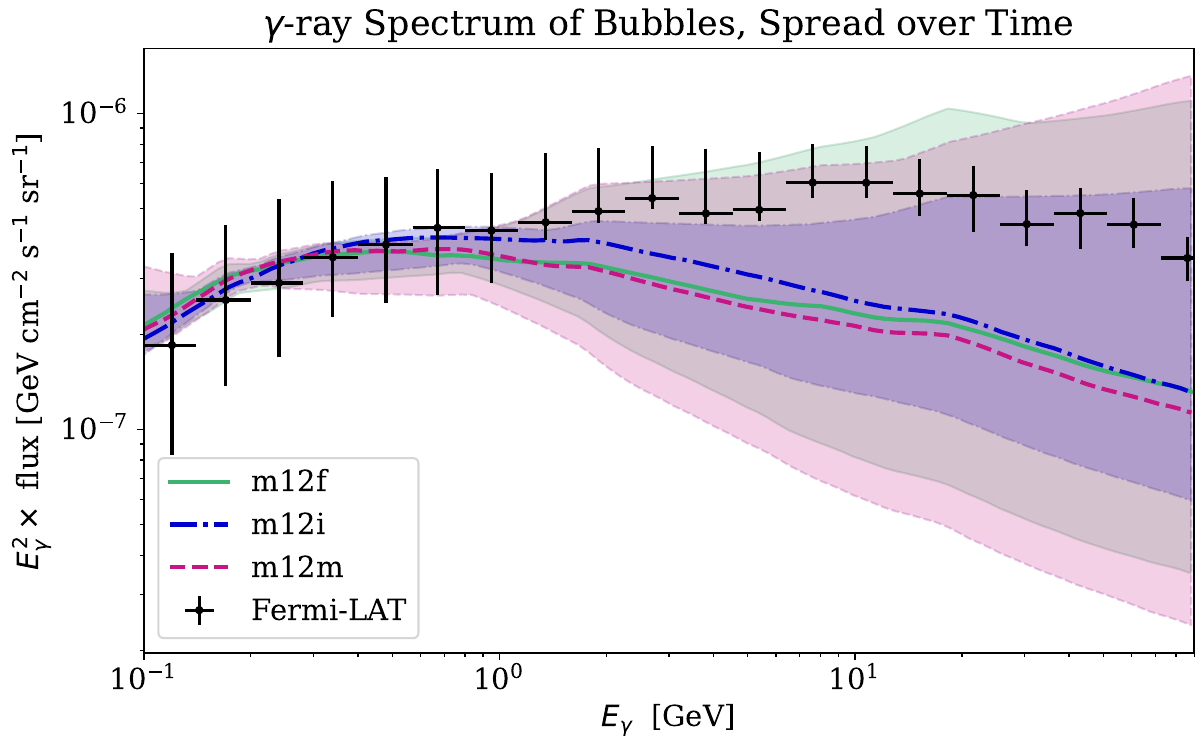}
    \includegraphics[width=\columnwidth]{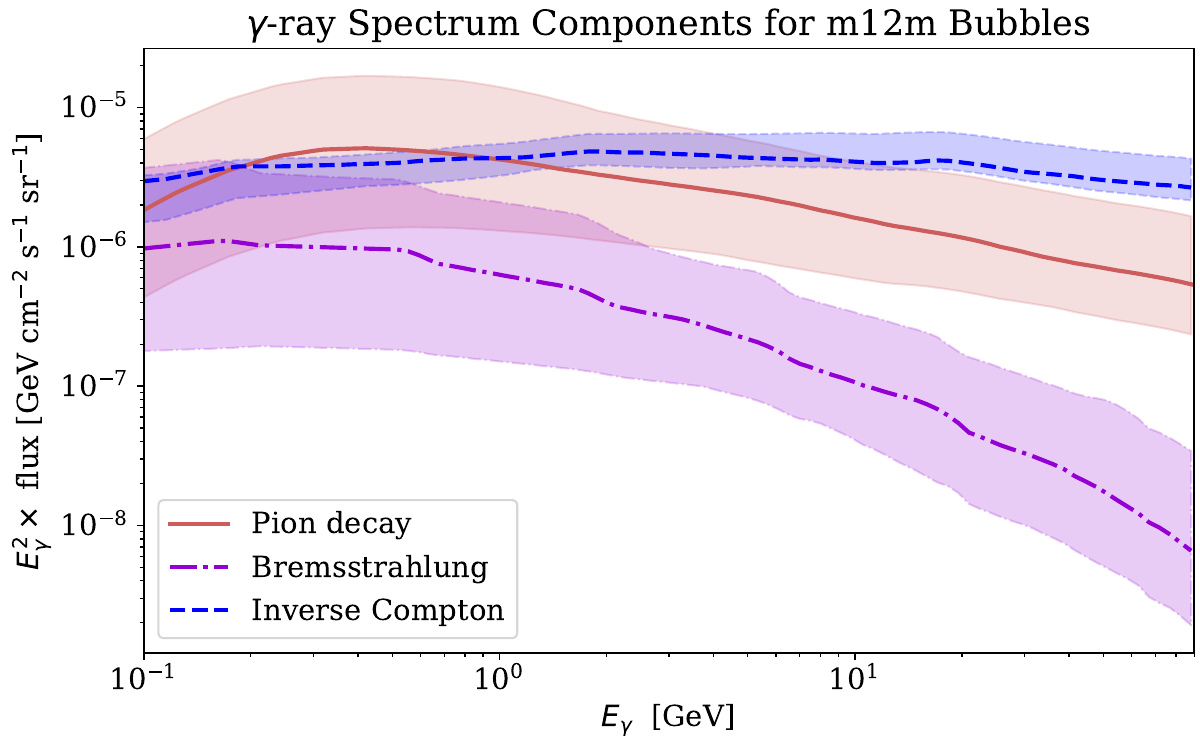}
    \caption{\textbf{Top}: the spread of $\gamma$-ray spectra from all CR interactions across all snapshots in the simulation, spaced between 460 Myr and 0 Myr in lookback time. A global renormalization to the spectrum amplitude has been applied to compare the spectra from simulations to each other and to the Fermi-LAT measurement. While the spectra for galaxies \textbf{m12f} and \textbf{m12m} is consistent with Fermi-LAT, \textbf{m12i} has a steeper tail at all times, due to high loss rates for CR leptons in the galactic center. \textbf{Bottom}: the components of the $\gamma$-ray spectra for the bubbles in galaxy \textbf{m12m}. $\gamma$-ray emission is produced by $\Pi_0$ decay, relativistic Bremsstrahlung, and inverse Compton scattering. The flux from inverse Compton is significantly less variable over time than the flux from $\Pi_0$ decay and relativistic Bremsstrahlung, both of which are proportional to gas density.}
    \label{fig:bubble_spec}
\end{figure}

\begin{figure*}
    \centering
    \includegraphics[width=0.49\linewidth]{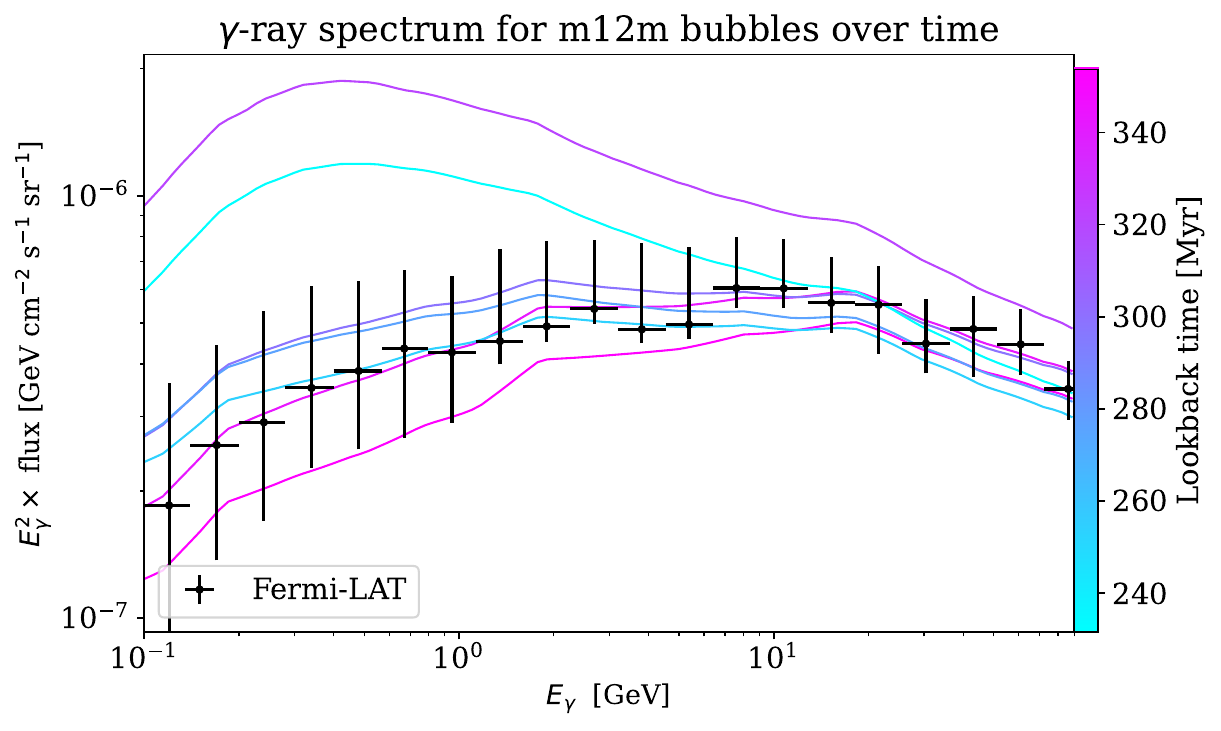}
    \includegraphics[width=0.49\linewidth]{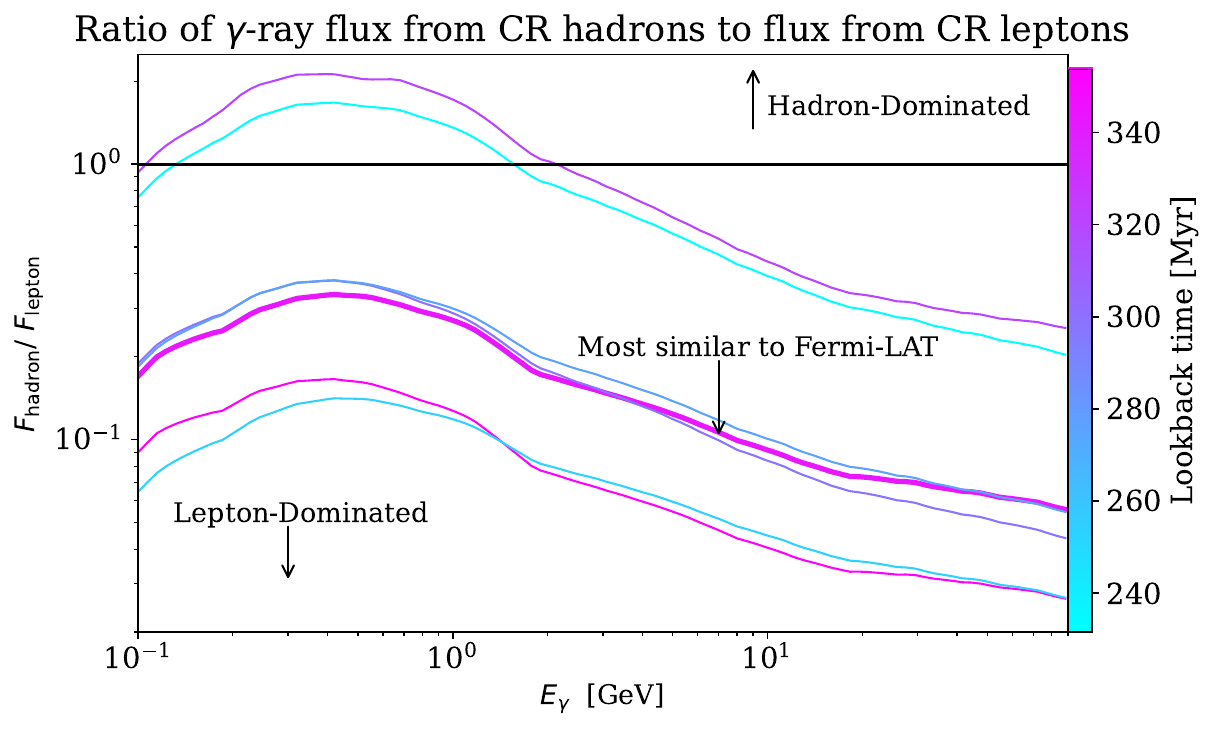}
    \caption{\textbf{Left}: The $\gamma$-ray spectrum for the bubbles in galaxy \textbf{m12m} at select snapshots between 340 and 240 Myr in lookback time. At 340 Myr, the shape of the spectrum is in good agreement with the measured spectrum of the Fermi bubbles. At 320 Myr, an inflow of gas causes a sharp increase in the spectrum, as well as a more peaked shape characteristic of hadronic spectra. After the gas settles, the spectra become flatter, until a cluster of supernovae go off at 240 Myr. \textbf{Right}: The ratio of $\gamma$-ray flux from $\Pi_0$ decay to flux from ICS for the same snapshots. The snapshot where the $\gamma$-ray spectrum is most similar in shape to the measured Fermi-LAT spectrum is indicated by the thick magenta line. Spectra corresponding to gas outflows/ inflows are more hadronic, while those more consistent with the Fermi-LAT measurement are lepton-dominated. However, there are times at which the $\gamma$-ray spectrum for the bubbles are more leptonic than the measured Fermi-LAT spectrum.}
    \label{fig:spec_ev}
\end{figure*}

The bubbles in galaxies \textbf{m12f} and \textbf{m12m} are brighter and have more structure than those in \textbf{m12i}. The fainter bubbles in \textbf{m12i} are likely attributable to the morphology of the galaxy, which features a strong magnetic field at the galactic center, and a warped inner disk that is nearly perpendicular to the outer disk. The strong magnetic field results in high CR lepton loss rates via synchrotron radiation, so fewer high-energy CR leptons escape the galactic center \cite{sands2025galacticcentergammarayemission}. In contrast, \textbf{m12f} and \textbf{m12m} have more typical disk structure and weaker magnetic fields, allowing CR leptons to stream into the halo.

Both the amplitude and the shape of the $\gamma$-ray spectrum of the bubbles vary by order-of-magnitude amounts over the 460 Myr span of the simulations. The amplitude of the $\gamma$-ray spectrum is typically 10-100 times higher than the spectrum measured by Fermi-LAT, due to the comparatively higher rate of star formation (and thus supernovae) relative to the MW in the simulations. However, the shape of the $\gamma$-ray spectrum, while highly variable over time, is consistent with the Fermi-LAT measurement (Fig. \ref{fig:bubble_spec}, top). The flux from ICS is relatively constant over time, while flux from $\Pi_0$ decay and relativistic Bremsstrahlung vary significantly (Fig. \ref{fig:bubble_spec}, bottom); the greater variability in the latter two processes is due to order-of-magnitude variations in gas density over time, while the cosmic ray flux and radiation background remain comparatively steady. The flux from relativistic Bremsstrahlung is typically subdominant to flux from both ICS and $\Pi_0$ decay. When the $\gamma$-ray flux from ICS exceeds the flux from $\Pi_0$ decay from $0.1 - 100$ GeV, the $\gamma$-ray spectrum is flatter, and closer in shape to the spectrum measured by Fermi-LAT. When the flux from $\Pi_0$ decay exceeds the flux from ICS, then the $\gamma$-ray spectrum has a pronounced peak at $E_{\gamma}\sim 0.3$ GeV, with a steeper tail. \\

\textbf{Growth and evolution of bubbles}: As there are no AGN in these simulations, the $\gamma$-ray bubbles are sourced by star formation, supernova feedback, and stellar winds, leading to a self-regulating ``galactic fountain" \cite{1980ApJ...236..577B,2008MNRAS.386..935F, 2015MNRAS.451.4223M, 2017ASSL..430..323F}. In all three simulated galaxies, there is a relatively small burst of star formation between 300 and 500 Myr ago, blowing gas and CRs out of the galactic center. Over a span of $\sim 100$ Myr, this gas falls back into the galactic disk and interacts with CR hadrons and leptons. Additional gas is supplied by supernova-driven outflows over the subsequent duration of the simulation. Figure \ref{fig:spec_ev} shows the $\gamma$-ray spectra for the bubbles in galaxy \textbf{m12m} at selected snapshots over this period of time. At 340 Myr, the spectrum is lepton-dominated and in good agreement with the Fermi-LAT measurement. At 320 Myr, infalling gas interacts with the CRs, resulting in greater $\gamma$-ray flux from $\Pi_0$ decay and a higher-amplitude, hadron-dominated spectrum. Eventually, the gas settles back into the disk, and the $\gamma$-ray spectrum of the bubbles becomes more leptonic. As this occurs, both the infalling gas and the compression of gas already in the disk from previous supernovae trigger further star formation \cite{2020MNRAS.493.4315M}. Finally, another cluster of supernovae goes off at 240 Myr, causing the $\gamma$-ray spectrum to be more hadronic once again. 

The most well-defined bubble structures form when there is little gas present from either inflows or outflows. The $\gamma$-ray spectrum calculated at such snapshots is predominantly leptonic, dominated by flux from ICS (Fig. \ref{fig:spec_ev}, right). The time the bubbles take to go through a ``cycle" in which an ICS-dominated bubble is inflated and then deflated by CRs interacting with an influx of gas is $\sim 50 - 100$ Myr. In galaxies \textbf{m12f} and \textbf{m12m}, several of these cycles occur, with oscillations between ICS-dominated and hadron-dominated $\gamma$-ray spectra for the duration of the simulation. In contrast, the $\gamma$-ray spectrum of the bubbles in galaxy \textbf{m12i}, which has a higher, relatively constant star formation rate for most of the duration of the simulation, become increasingly hadronic over time, particularly as strong magnetic fields in the galactic center deplete the CR leptons \cite{sands2025galacticcentergammarayemission}.

\begin{figure*}
    \centering
    \includegraphics[width=\linewidth]{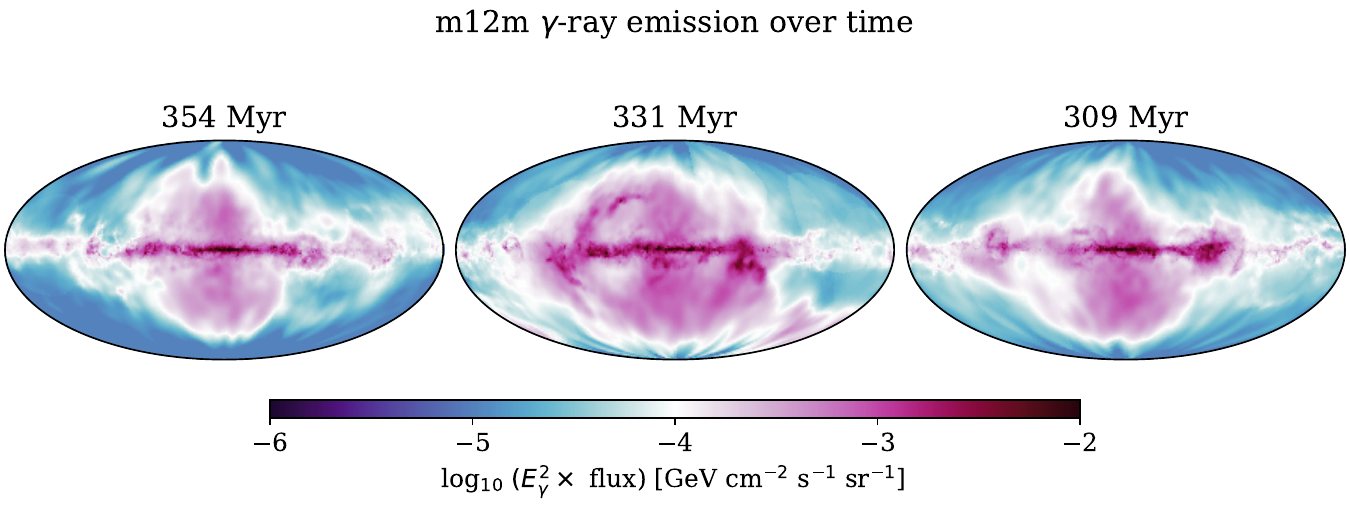}
    \includegraphics[width=\linewidth]{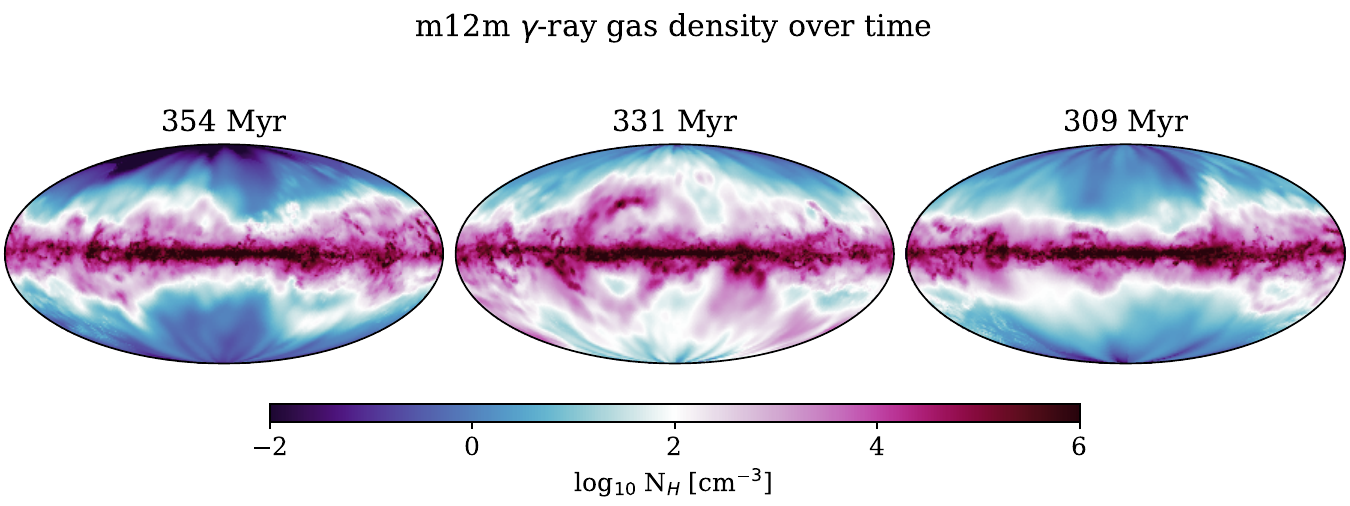}
    \caption{$\gamma$-ray emission (top) and gas density (bottom) for galaxy \textbf{m12m} over a 45 Myr span. At 354 Myr, a clear bi-lobed $\gamma$-ray bubble is visible. At 331 Myr, the bubbles become brighter and less ordered as gas previously ejected by supernova feedback falls back into the disk. By 309 Myr, the gas has re-settled in the disk, and the distinct bi-lobed shape of the bubbles is visible once more.}
    \label{fig:bubblegas}
\end{figure*}

\begin{figure*}[t]
    \centering
    \includegraphics[width=0.49\linewidth]{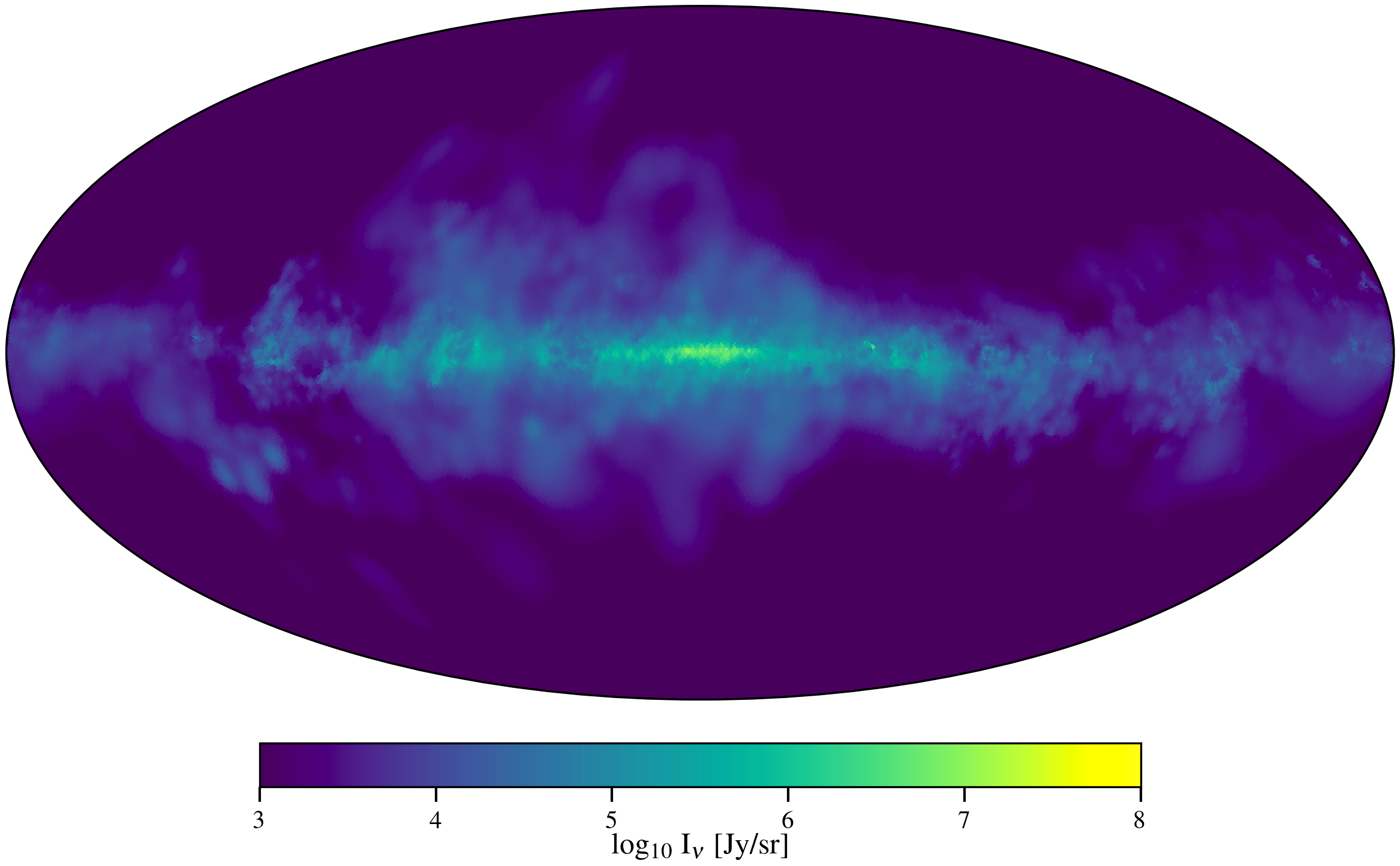}
    \includegraphics[width=0.49\linewidth]{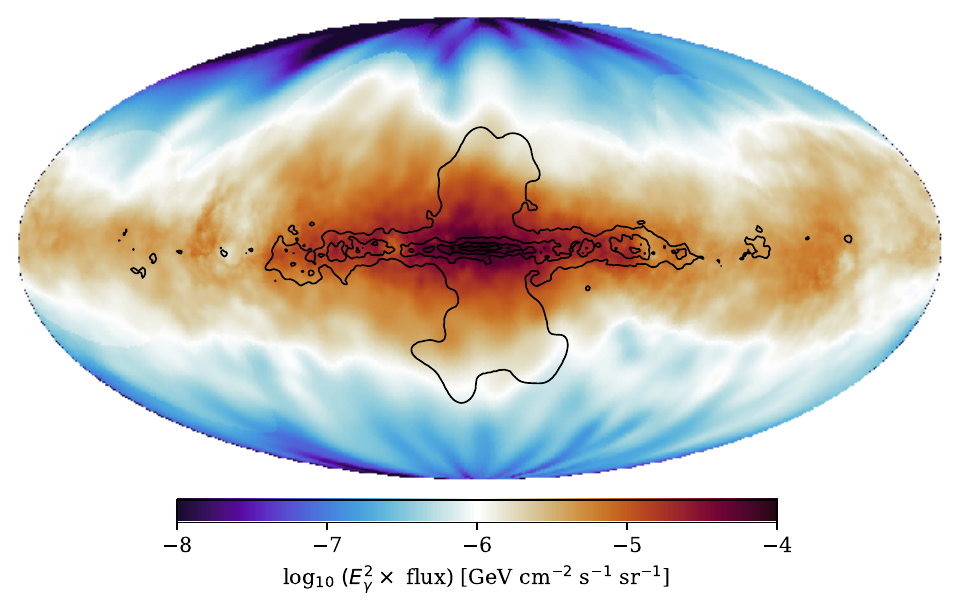}
    \caption{ Observable features in other wavelengths related to the $\gamma$-ray bubbles in galaxy \textbf{m12m}. \textbf{Top}: Synchrotron radio emission at 20 GHz from CR leptons in galaxy \textbf{m12m}. The flux is roughly comparable to the observed flux of the WMAP haze, and there is structure similar to observed radio loops and spurs in the MW \cite{2004ApJ...614..186F} \textbf{Bottom}: soft x-ray (0.1-2.4 keV) surface brightness from inverse Compton scattering of CR leptons in galaxy \textbf{m12m}. Contours for $\gamma$-ray flux are shown in the black lines, with each contour separated by $10^{0.5} \, \rm{GeV \, cm^{-2} \, s^{-1} \, sr^{-1}}$. The edges of the $\gamma$-ray bubbles and x-ray halo are roughly coincident. Both of these observable features arise from CR lepton interactions (for synchrotron, with magnetic fields, and for ICS, with radiation), and are bright enough to potentially be detected by current and future surveys.}
    \label{fig:multiWL}
\end{figure*}

\textbf{Multi-wavelength observations: x-ray and synchrotron radio}: The Fermi bubbles have associated observable features in other wavelengths, including the eROSITA bubbles (soft x-ray) \cite{2020Natur.588..227P}, the WMAP haze (microwave and radio) \cite{2004ApJ...614..186F, 2008ApJ...680.1222D, 2013A&A...554A.139P}, and Loop 1 and the South Polar Spur (radio) \cite{2015MNRAS.452..656V, 2016A&A...594A..25P}. We find that the bubbles in our simulations have counterparts in x-ray and synchrotron radio emission (Fig. \ref{fig:multiWL}). While the dominant morphology of the emission at these wavelengths is a diffuse halo, some geometry of the bubbles is present  (particularly in southern hemisphere in the soft x-ray). The edge of the $\gamma$-ray bubbles coincides with the edge of the x-ray halo. Both the synchrotron radio and the x-ray emission are bright enough to be feasibly detected by current and next-generation radio and x-ray telescopes, respectively. We leave further exploration of multi-wavelength observables for galactic outflows in spectral MHD-CR simulations to future work. 

\textbf{Discussion}: This work demonstrates that feedback from supernovae and stellar winds can give rise to Fermi bubble-like $\gamma$-ray outflows, without invoking AGN feedback or beyond standard model physics. Because $\gamma$-ray outflows and their counterparts in other wavelengths occur in all three simulated galaxies in our sample, it is plausible that these features are very common, but difficult to detect in galaxies beyond the MW and the Local Group.  $\gamma$-ray bubbles that form by this mechanism are in best agreement with the observed Fermi bubble $\gamma$-ray spectrum when most of the $\gamma$-ray flux is produced via ICS, with a $\sim 10-20 \%$ contribution from $\Pi_0$ decay and relativistic Bremsstrahlung. These structures are highly dynamical, evolving on the scale of tens of Myrs, over which their composition can change dramatically.  

These results do not rule out the possibility that AGN-accelerated cosmic rays contribute to the formation of such structures. Rather, by including AGN feedback in future simulations, we can leverage the Fermi bubble observations as a constraint on models of CR transport and AGN physics. There are features of the Fermi and eROSITA bubbles (e.g., the hard edge to the bubble surface brightness, shape of the x-ray emission) that are not reproduced in some of these simulations. If such features occur in simulations with CR injection from both AGN feedback and star formation, or from AGN alone, they could help discriminate between these scenarios. Additionally, forward-modeling of $\gamma$-ray and x-ray observables in the simulation can provide insight into whether the hard edge of the Fermi bubbles arises from physics (e.g., a shock) or observational effects, such as geometry or modeling of foregrounds and backgrounds.

Beyond the formation of the Fermi bubbles, this work has important implications for and connections to other facets of galaxy formation and astro-particle physics. If $\gamma$-ray bubbles typically form via stellar feedback-driven outflows, then the existence of the Fermi bubbles may indicate a relatively recent starburst (within last $\sim$ 100 Myr) in the MW's galactic center. There are numerous observational and theoretical challenges for measuring the star formation history of the MW's central regions \cite{2023ASPC..534...83H}. Simulations have indicated that star formation in the galactic center of MW-like galaxies is typically bursty, with feedback-driven intervals in star formation rate of $\sim 10-100$ Myr \cite{2017MNRAS.467.2301T, 2021ApJ...908L..31O}. Historically, star formation in this region was thought to be quasi-continuous over the last 10 Gyr, but recent observations suggest that there was a burst of star formation in the MW's nuclear stellar disk in the last 30 Myr \cite{Nogueras_Lara_2019}. Future observational surveys of young stars in the galactic center will help to constrain the star formation history in the galactic center, and can test stellar feedback as a formation mechanism for the Fermi bubbles. 

The processes that gives rise to the $\gamma$-ray bubbles in these simulations directly affect the CGM, where CR transport is both less constrained and has greater impact on galaxy formation than in the ISM. The $\gamma$-ray bubbles form through a cycle of galactic outflows and inflows that has hadron-dominated and lepton-dominated phases; the hadron-dominated phases are typically associated with galactic fountain flows. During the lepton-dominated phases, the $\gamma$-ray bubbles are expected to have counterparts in diffuse emission in other wavelengths, such as synchrotron radio and soft x-ray from ICS. Although detecting diffuse $\gamma$-ray emission around other galaxies is difficult due to the faintness of these features, there are better prospects for observing multi-wavelength counterparts with next-generation observatories . Detecting the these multi-wavelength signatures of outflows in a larger sample of galaxies could determine how much time a MW-like galaxy spends in each phase and inform our understanding of the inner CGM.  In addition, these results support a growing body of work that indicates that, rather than all CRs free-streaming into the CGM and beyond, a sizable fraction must inhabit a large ( $\sim$ 10 kpc) CR ``scattering'' halo-- in these simulations, the CRs that produce $\gamma$-ray bubbles persist at radii of $\sim$ 5-10 kpc above the galactic disk over tens of Myrs. CR leptons that escape the smaller scattering halo can diffuse into the outer CGM, where, by inverse Compton scattering off of CMB photons, they can produce non-thermal x-ray emission at large radii ($\sim$ 100 kpc) \cite{hopkins2025cosmicraysmasqueradinghot}. 

Uncovering the origins of the Fermi bubbles is a complex problem due to the multitude of astrophysical processes and scales involved. The ability to simulate features similar to the Fermi bubbles in full cosmological context and forward-model their emission from $\gamma$-ray to radio frequencies will enable new tests of their potential formation mechanisms. 

%\textit{The next generation of x-ray and $\gamma$-ray observatories may be able to shed light on how frequently these outflows occur. Such outflows and their multi-wavelength signatures can serve as powerful observational constraints on CR transport, and lead to new insights about the history of activity in the Milky Way's galactic center.} 

%GR bubbles form through a cycle that has hadron dominated and lepton dominated phases; hadron dominated correspond to galactic fountain. Lepton dominated, we should expect to see radio and x-ray counterparts. Hadron dominated, shouldn't see anything. Could be used to determine frequency/ how much time a galaxy spends in each phase, but difficult because gamma-ray emission is faint. 

%\clearpage
%\newpage

\bibliography{apssamp}% Produces the bibliography via BibTeX.

\end{document}